\documentclass[12pt]{article}
\usepackage{epsfig}
\def \k{\kappa}

\title{Effective Drift and Diffusivity in non-Gaussian Random Gradient Flows}
\author{I T Drummond, R R Horgan and C A da Silva Santos\\
Department of Applied Mathematics and Theoretical Physics\\
University of Cambridge\\
Silver St\\
Cambridge, England CB3 9EW}
\begin{document}
\maketitle
\begin{abstract}

We study the long-range effective drift and diffusivity of a particle 
in a random medium moving subject to a given molecular diffusivity
and a local drift. The local drift models the effect of a random 
electrostatic field on a neutral but polarizable molecule. Although
the electrostatic field is assumed to obey Gaussian statistics the
induced statistics of the drift velocity field are non-Gaussian.

We show that a four-loop perturbation theory calculation of the effective diffusivity 
is in rather good agreement with the outcome of a numerical simulation for 
a reasonable range of the disorder parameter. We also measure the effective drift 
in our simulation 
and confirm the validity 
of the ``Einstein relation'' that expresses the equality of the renormalization
factors, induced by the random medium, for the effective 
drift and effective diffusivity, relative to their molecular values. 
The Einstein relation has previously only been confirmed for Gaussian
random drift fields. The simulation result, for our non-Gaussian drift model,
is consistent with a previous theoretical analysis 
showing the  Einstein relation should remain true, independently of the precise character
of the statistics of the drift velocity field.
\end{abstract}
\vspace{4cm}
DAMTP-97-10
\newpage

\section{Introduction}

The advective diffusion of passive scalar fields in random environments 
has been extensively studied by both analytical and numerical techniques, 
with particular emphasis on the computation of effective parameters for a diffusion process
that combines molecular diffusion with a drift term that depends linearly on the
gradient of a random scalar field \cite{Bouch} - \cite{DDH5}.

The problem is well studied and understood in the case of transport by a 
gradient velocity field which exhibits {\it Gaussian} statistics. In isotropic
systems a renormalization group approach can be shown to give exact results
in one and two dimensions \cite{Bouch,Krav1,Krav2}. Somewhat surprisingly the same approach works 
extremely well for the isotropic problem in three dimensions \cite{Deem,DeanTh,DDH1,DDH2}, 
though the situation is 
less clear in the absence of isotropy \cite{DDH3,DDH4}. 

One ingredient in the success of the 
renormalization group method is that, at each stage of the calculation, 
it respects the Einstein relation that guarantees the equality of the renormalization factors
of the effective diffusivity, $\kappa_e$, and drift, $\lambda_e$, 
parameters relative to their molecular values, $\kappa_0$ and $\lambda_0$ repectively.
That is
\begin{equation}
\frac{\kappa_e}{\kappa_0}=\frac{\lambda_e}{\lambda_0}~~.
\end{equation}
This result does not hold in the, somewhat arbitrary, Hartree-Fock resummation procedure
which turns out to be even less accurate than simple low order perturbation theory which
also breaks down for strong disorder.

It turns out that the Einstein relation holds in a wide range of circumstances
including ones where the diffusivity and drift coefficient have non-trivial 
tensorial structure \cite{DDH5}. The only requirement is that the two tensors are linearly dependent.
Thus if at the molecular level we have $\lambda_{0~ij}=\tau\kappa_{0~ij}$ then we will
find for the macroscopic effective parameters that $\lambda_{e~ij}=\tau\kappa_{e~ij}$~.
The isotropic situation is an example of this. The reason that the proportionality
factor survives renormalization can be traced back to the existence of a finite sample
equilibrium distribution with a vanishing micro-current. In equilibrium statistical
mechanics this vanishing is guaranteed which explains the emergence of the
Einstein relation in this context. In the quenched models with which we are concerned
the Einstein relation is not guaranteed except in the special but important circumstances
indicated above. Given these conditions of proportionality of the molecular
tensors however, the Einstein relation can be shown to hold sample by sample.
It follows that the relation will hold {\it independent} of the precise nature of the
statistics of the gradient velocity field. This theoretical prediction \cite{DDH5}
lends considerable interest to the results of the numerical simulation of the
physically motivated model that we study in this paper, 
in which the gradient velocity field does not
have Gaussian statistics.

A physical realization of the {\it Gaussian} case  
is the diffusion of an ion in a sea of fixed, 
disordered charges that give rise to an electrostatic potential. Such a situation can be 
created by embedding fixed ions in the rather accommodating structure
of a zeolite matrix. It is not unreasonable 
to treat the resulting electrostatic field as having Gaussian statistics 
at length scales somewhat larger than
the molecular level but very short compared to macroscopic scales. 
A related problem, particularly relevant to
applications in chemical processes, concerns the diffusion of a {\it neutral} molecule 
in such a medium. The case of the diffusion of benzene in zeolites with random
ionic inclusions has been studied experimentally and has 
important technological applications \cite{gladden}.
In this situation the force on the molecule arises because of the interaction of its
induced dipole moment with the electrostatic field produced by the included ions. The result
is a gradient velocity field with {\it non}-Gaussian statistics. For simplicity 
we take it for granted that all tensorial effects are absent.

We assume that the dipole moment $P_i$ of the diffusing particle is proportional 
to the ambient electric field $E_i$ thus
\begin{equation}
P_i=\mu E_i~~.
\label{polarize}
\end{equation}
The force on the molecule is 
\begin{equation}
F_i=P_j\partial_jE_i~~,
\label{force}
\end{equation}
and since for an electrostatic field $\partial_jE_i=\partial_iE_j$~, it follows that
\begin{equation}
F_i=\mu E_j\partial_iE_j=\frac{1}{2}\mu\partial_iE_j^2~~.
\end{equation}
If we introduce the electrostatic potential $\phi({\bf x})$ then the
force on the molecule becomes
\begin{equation}
F_i=\frac{1}{2}\mu\partial_i(\nabla\phi({\bf x}))^2~~.
\end{equation}
If we assume that the drift of the molecule is given by
\begin{equation}
u_i({\bf x})=\nu F_i~~,
\end{equation}
then, absorbing all constants into an overall parameter $\lambda_{0}$, we have
\begin{equation}
u_i({\bf x})=\frac{1}{2}\lambda_{0} \partial_i(\nabla\phi({\bf x}))^2~~.
\end{equation}
It is still reasonable to model the statistical properties of the electrostatic
potential $\phi({\bf x})$ by a Gaussian field. However because of the quadratic
relationship between $u_i({\bf x})$ and $\phi({\bf x})$ the
statistics of the drift velocity are {\it not} Gaussian in this model. The higher
cumulants of the velocity field beyond the second do not vanish. Indeed it is very
clear that the third and other odd order cumulants exist. This means that the 
problem is not symmetrical under the change $\lambda_{0}\rightarrow -\lambda_{0}$~. For this
reason we explore both positive and negative values of $\lambda_{0}$ even though the
physical derivation of the model suggests that $\lambda_{0}>0$~.
For the purposes of the simulation we take the opportunity to absorb the
normalization of $\phi({\bf x})$ into the definition of $\lambda_0$ and require
\begin{equation}
\langle (\phi({\bf x}))^2\rangle=1~~.
\end{equation}

The problem we address then, is the evaluation of the long-range effective
parameters in terms of the (non-Gaussian) statistical properties of the 
random flow. Such a model presents its own new technical difficulties. Calculational schemes
such as Self-Consistent Perturbation Theory or the Renormalization Group (RG), which proved so 
successful in Gaussian problem, turn out to be hard to apply to 
this particular problem. The barriers to a straightforward application of these 
more sophisticated perturbation methods 
arise mainly because of the increased complexity of the vertex structure
at low wave number that is quickly revealed by perturbation theory.
In addition the lowest order correction to the propagator begins at two loops rather
than one loop as in the Gaussian case. 
In this paper we confine ourselves to computing the standard perturbation
expansion for the effective diffusivity to four-loop order. As will become clear
the results are consistent with the outcome of our numerical simulation of the model
over a significant range of values for the drift parameter. 

\section*{2 Perturbation Theory}

The diffusion equation in which we are interested has the form
\begin{equation}
\frac{\partial P({\bf x},t)}{\partial t} 
            = \nabla (\k_{0} \nabla P({\bf x},t)-{\bf u(x)}P({\bf x},t))~~.
\end{equation}
Here, $P({\bf x},t)$ is the probability density of a particle moving 
according to the equation
\begin{equation}
{\bf {\dot{x}}} = {\bf u(x)} + {\bf w}(t)~~,
\label{sde}
\end{equation}
where ${\bf w}(t)$ is a white noise term that satisfies
\begin{equation}
\langle w_{i}(t)\rangle = 0\;\;\; 
    \mbox{and} \;\;\;\langle w_{i}(t)w_{j}(t')\rangle = 2\k_{0}\delta _{ij}\delta (t-t')~~.
\end{equation}
The flow field, $\bf u(x)$, is taken to be time independent  and is the 
gradient of a scalar field
\begin{equation}
{\bf u(x)} = \lambda _{0}\nabla \psi (\bf x)~~,
\end{equation}
but, because the flow originates in the interaction of the induced dipole moment 
of the diffusing particle with the electrostatic field, 
$\psi(\bf x)$ does not exhibit Gaussian statistics. 
In fact, as indicated above, we have
\begin{equation}
\psi ({\bf x}) = \frac{1}{2}(\nabla \phi ({\bf x}))^{2}~~,
\end{equation}
where $\phi (\bf x)$ is a homogeneous Gaussian random field characterized by 
the disorder averages
\begin{equation}
\langle\phi ({\bf x}) \rangle =0\;\;\;\mbox{and}\;\;\;
   \langle\phi ({\bf x})\phi ({\bf y}) \rangle = \Delta ({\bf x-y})~~.
\end{equation}
For simplicity we take the disorder to be isotropic, that is,  
$\Delta = \Delta (|{\bf x|})$.

\begin{figure}[htb]
\begin{center}
\epsfig{file=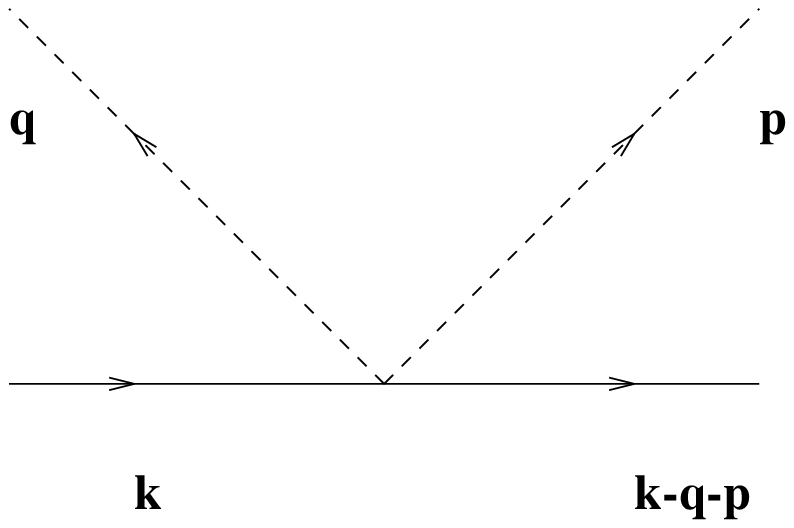,height=30mm}
\end{center}
\caption{\label{f1}Vertex diagram.}
\end{figure}

The perturbative approach to solving equation (1) is well known 
\cite{Bouch,Krav1,Krav2,DDH1,Phyth,DrDuH,DrH} and we only summarize here the necessary results.
Since we are interested in the effective parameters governing the 
evolution of the distribution $P({\bf x},t)$, we study the related 
static Green function, $G(\bf x)$, which satifies
\begin{equation}
\k_{0}\nabla ^{2} G({\bf x})-\nabla ({\bf u(x)}G({\bf x})) = -\delta({\bf x})~~.
\end{equation}
A perturbation series in the coupling $\lambda _{0}$ for $\tilde{G}({\bf k})$ 
can be generated by iterating the formal solution to equation (7) in Fourier space: 
\begin{equation}
\tilde{G}({\bf k}) = \frac{1}{\k_{0}{\bf k}^{2}}
-\frac{\lambda_{0}}{\k_{0}{\bf k}^{2}}
\int\frac{d^{3}{\bf q}}{(2\pi)^{3}}\frac{d^{3}{\bf p}}{(2\pi)^{3}}\,
\tilde{G}({\bf k-q-p})\,\tilde{\phi}({\bf q})\,\tilde{\phi}({\bf p})\,
 \frac{{\bf q\cdot p\,k\cdot (q+p)}}{2}~~.
\end{equation}
The Green function averaged over the velocity ensemble, $\langle\tilde{G}({\bf k})\rangle$, 
can be written as
\begin{equation}
\langle\tilde{G}({\bf k})\rangle = \frac{1}{\k_{0}{\bf k}^{2}-\Sigma({\bf k})}~~,
\end{equation}
where the averaging over the velocity ensemble is done using Wick's theorem 
to give a diagrammatic expansion and $\Sigma({\bf k})$ is the summation of 
one particle-irreducible diagrams. The expected asymptotic behaviour of 
the diffusion process at large distances and times implies that the small 
$\bf k$ behaviour of $\langle\tilde{G}({\bf k})\rangle$ is given by
\begin{equation}
\k_{e} = \k_{0} - \frac{d}{d{\bf k}^{2}}\Sigma({\bf k})|_{{\bf k}=0}~~,
\end{equation}
where $\k_{e}$ is the effective diffusivity.
The Feynman rules for the diagrammatic expansion are as follows:
\begin{enumerate}
\item Wave vector is conserved at each vertex;
\item Each full line carries a factor $1/\k_{0}{\bf k}^{2}$;
\item Wave vector is integrated around closed loops with a factor $d^3{\bf q}/(2\pi)^{3}$;
\item Each vertex, whose diagrammatic representation is shown in figure \ref{f1}, 
carries a factor  $-\lambda_{0}\,{\bf q\cdot p\,k\cdot (q+p)}$;
\item Each internal dashed line carries a factor $\tilde{\Delta}({\bf q})$;
\item Each diagram must be divided by the usual symmetry factor.  
\end{enumerate}
In what follows, we use the explicit spectrum
\begin{equation}
\tilde{\Delta}({\bf q})=\frac{(2\pi)^{3/2}}{k_{0}}\,e^{-{\bf q}^{2}/2k_{0}^{2}}~~.
\end{equation}
The normalization is chosen so that $\langle(\phi({\bf x}))^{2}\rangle = 1$. 
In our numerical calculations, we set $k_{0} = 1$.

There is no one loop correction to the propagator. As mentioned before, 
this together with the fact that {\it new} vertices are generated when correcting 
the primitive one makes it extremely delicate to implement other perturbative 
schemes such as Self-Consistent or RG methods. Therefore, we concentrate on a 
straightfoward perturbation theory calculation which has to be carried out to 
at least four-loop order to get a sensible outcome. The formal manipulations of which we 
make use are basically the same as those utilized in \cite{DeanTh,DDH1,DDH2} 
and we only elucidate 
the more elaborate steps and state the main results.

\begin{figure}[htb]
\begin{center}
\epsfig{file=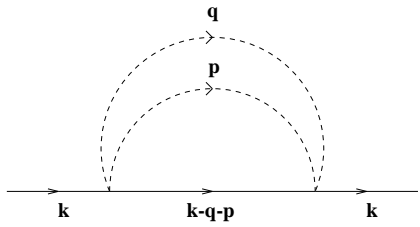,height=30mm}
\end{center}
\caption{\label{f2}Two-loop contribution to $\Sigma(\bf k)$.}
\end{figure}

The two-loop contribution to $\Sigma({\bf k})$ is associated with the diagram 
in figure \ref{f2}. According to the above Feynman rules it is
\begin{equation}
\Sigma^{(2)}({\bf k}) = -\frac{1}{2}\frac{\lambda_{0}^{2}}{\k_{0}}
\int\frac{d^{3}{\bf q}}{(2\pi)^{3}}\frac{d^{3}{\bf p}}{(2\pi)^{3}}\,
\tilde{\Delta}({\bf q})\,\tilde{\Delta}({\bf p})\,
\frac{{\bf (p\cdot q)}^{2}\,{\bf k\cdot (p+q)\,(k-q-p)\cdot (q+p)}}{{\bf (k-q-p)}^{2}}~~,
\end{equation}
and it can be easily computed to $O({\bf k}^{2})$ with the result
\begin{equation}
{\Sigma}^{(2)}({\bf k}) = \frac{1}{2}\,\frac{\lambda_{0}^{2}}{\k_{0}}\,{\bf k}^{2}~~.
\end{equation}

\begin{figure}[htb]
\begin{center}
\epsfig{file=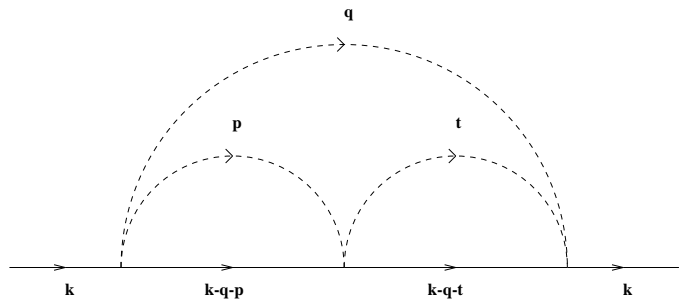,height=40mm}
\end{center}
\caption{\label{f3}Three-loop contribution to $\Sigma(\bf k)$.}
\end{figure}

In order to encounter deviations from Gaussian behaviour we must calculate beyond second order 
and consider the three-point correlator of the velocity field, that is, 
to include the first non-zero odd power in $\lambda_{0}$ in the perturbation expansion. 
The three-loop diagram contributing to $\Sigma({\bf k})$ is shown in figure \ref{f3} and yields
\begin{eqnarray}
{\Sigma}^{(3)}({\bf k}) & = & -\frac{\lambda_{0}^{3}}{\k_{0}^{2}}
\int\frac{d^{3}{\bf q}}{(2\pi)^{3}}\frac{d^{3}{\bf p}}{(2\pi)^{3}}\frac{d^{3}{\bf t}}{(2\pi)^{3}}\,
\tilde{\Delta}({\bf q})\,\tilde{\Delta}({\bf p})\,\tilde{\Delta}({\bf t})\,
{\bf q\cdot p\,p\cdot t\,q\cdot t}\nonumber \\
                &   & \cdot\,\frac{{\bf k\cdot (q+p)\,(k-q-p)\cdot (t-p)\,
(k-q-t)\cdot (q+t)}}{{\bf (k-q-p)}^{2}\,{\bf (k-q-t)}^{2}}~~.
\end{eqnarray}
The numerator can be re-written using the identity
\begin{equation}
{\bf (k-q-t)\cdot (-q-t)} = {\bf (k-q-t)}^{2}-{\bf k\cdot (k-q-t)}~~.
\label{identity}
\end{equation}
We then have
\begin{eqnarray}
\Sigma^{(3)}({\bf k}) & = & -\frac{\lambda_{0}^{3}}{\k_{0}^{2}}
\int\frac{d^{3}{\bf q}}{(2\pi)^{3}}\frac{d^{3}{\bf p}}{(2\pi)^{3}}\frac{d^{3}{\bf t}}{(2\pi)^{3}}\,
\tilde{\Delta}({\bf q})\,\tilde{\Delta}({\bf p})\,\tilde{\Delta}({\bf t})\,
{\bf q\cdot p\,p\cdot t\,q\cdot t}\nonumber \\
         &   & {\bf k\cdot (q+p)\,(k-q-p)\cdot (p-t)}\,\left\{\frac{1}{{\bf (k-q-p)}^{2}}\right.
\nonumber \\
         &   & \left.-\frac{\bf k\cdot (k-q-t)}{{\bf (k-q-p)}^{2}\,{\bf (k-q-t)}^{2}}\right\}~~.
\label{pert3}
\end{eqnarray}
The first term, which turns out to be the dominant one, can be performed analytically 
whereas the second one, much smaller, has to be done numerically to $O({\bf k}^{2})$. 
It is instructive to explain in some detail how to compute the analytical contribution to 
$\Sigma^{(3)}({\bf k})$. In the first term of equation (\ref{pert3}), the term odd in $\bf t$ 
integrates to zero and, moreover, it can be symmetrized for $\bf p$ and $\bf q$ to 
yield the result
\begin{eqnarray}
\Sigma_{an}^{(3)}({\bf k}) & = & -\frac{1}{2}\frac{\lambda_{0}^{3}}{\k_{0}^{2}}
\int\frac{d^{3}{\bf q}}{(2\pi)^{3}}\frac{d^{3}{\bf p}}{(2\pi)^2}\frac{d^{3}{\bf t}}{(2\pi)^2}\,
\tilde{\Delta}({\bf q})\,\tilde{\Delta}({\bf p})\,\tilde{\Delta}({\bf t})\,
{\bf q\cdot p\,p\cdot t\,q\cdot t}\nonumber \\
              &   & \cdot \,\frac{\bf k\cdot (q+p)\,(k-q-p)\cdot (q+p)}{{\bf (k-q-p)}^{2}}~~.
\end{eqnarray}
Using again equation (\ref{identity}) (for vectors $\bf p$ and $\bf q$) leads to
\begin{eqnarray}
\Sigma_{an}^{(3)}({\bf k}) & = & \frac{1}{2}\frac{\lambda_{0}^{3}}{\k_{0}^{2}}
\int\frac{d^{3}{\bf q}}{(2\pi)^{3}}\frac{d^{3}{\bf p}}{(2\pi)^{3}}\frac{d^{3}{\bf t}}{(2\pi)^{3}}\,
\tilde{\Delta}({\bf q})\,\tilde{\Delta}({\bf p})\,\tilde{\Delta}({\bf t})\,{\bf q\cdot p\,p\cdot t\,
q\cdot t}\,{\bf k\cdot (q+p)}\nonumber \\
              &   & \cdot\,\left\{1-\frac{\bf k\cdot (k-q-p)}{{\bf (k-q-p)}^{2}}\right\}~~.
\label{momid}
\end{eqnarray}
The first term in equation (\ref{momid}) integrates to zero. We need to evaluate the second 
term only to $O({\bf k}^{2})$. Because of the explicit factors of $\bf k$ in the 
integrand we can set ${\bf k}=0$ everywhere else to obtain
\begin{eqnarray}
\Sigma_{an}^{(3)}({\bf k}) & = & \frac{1}{2}\frac{\lambda_{0}^{3}}{\k_{0}^{2}}\int\frac{d^{3}{\bf q}}{(2\pi)^{3}}\frac{d^{3}{\bf p}}{(2\pi)^{3}}\frac{d^{3}{\bf t}}{(2\pi)^{3}}\,\tilde{\Delta}({\bf q})\,\tilde{\Delta}({\bf p})\,\tilde{\Delta}({\bf t})\,{\bf q\cdot p\,p\cdot t\,q\cdot t}\nonumber \\
                         &   & \cdot \frac{\bf k\cdot (q+p)\,k\cdot (q+p)}{{\bf (q+p)}^{2}}~~.
\end{eqnarray}
This is easily evaluated as
\begin{equation}
\Sigma_{an}^{(3)}({\bf k}) = \frac{1}{2}\frac{\lambda_{0}^{3}}{\k_{0}^{2}}\,
{\bf k}^{2}\,{\left[\frac{1}{3}\int\frac{d^{3}{\bf q}}{(2\pi)^{3}}\,{\bf q}^{2}\,
\tilde{\Delta}({\bf q})\right]}^{3} = \frac{1}{2}\frac{\lambda_{0}^{3}}{\k_{0}^{2}}\,{\bf k}^{2}~~.
\end{equation}
Combining the analytical and numerical pieces yields the following result
\begin{equation}
\Sigma^{(3)}({\bf k}) = \frac{\lambda_{0}^{3}}{\k_{0}^{2}}\,{\bf k}^{2}\,(\frac{1}{2}+0.030375)~~.
\end{equation}

\begin{figure}[htb]
\begin{center}
\epsfig{file=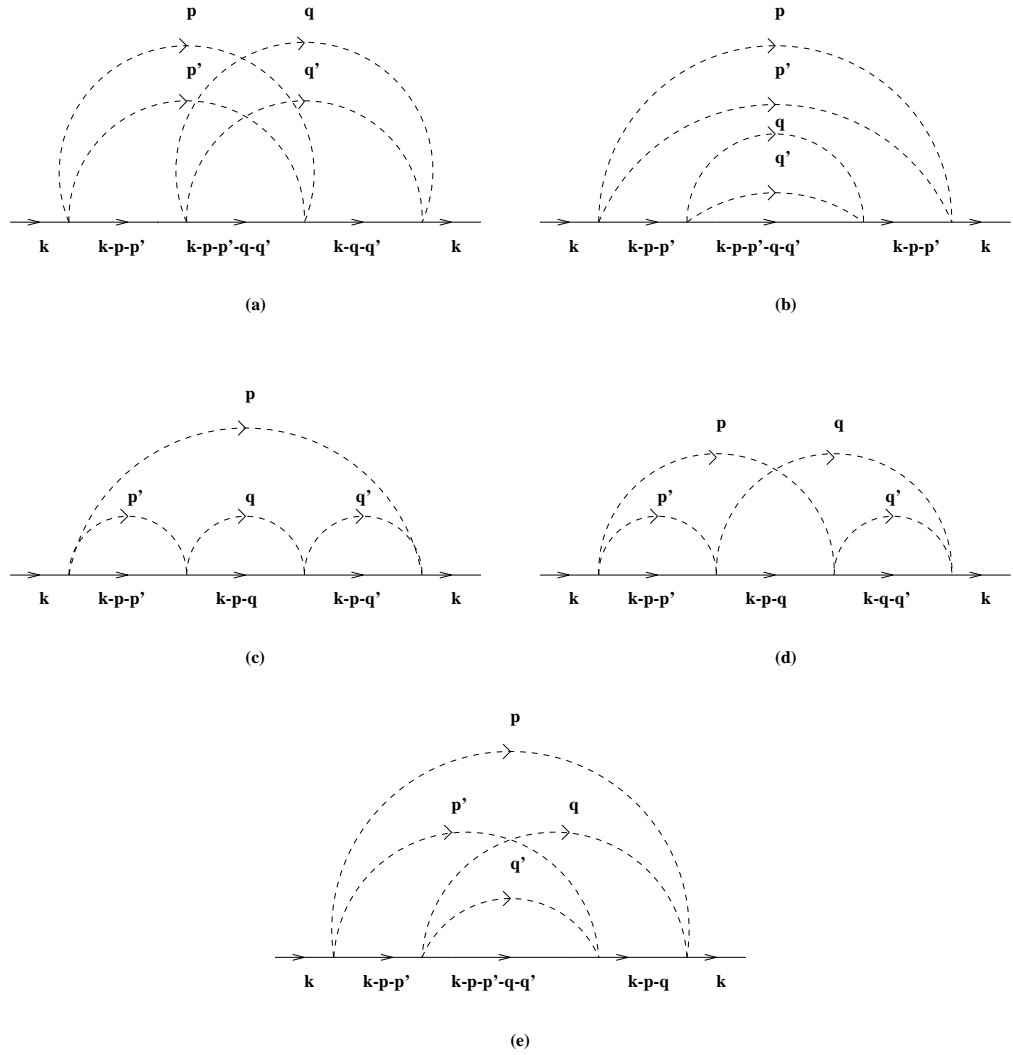,height=140mm}
\end{center}
\caption{\label{f4}Four-loop contributions to $\Sigma(\bf k)$.}
\end{figure}

\begin{figure}[htb]
\begin{center}
\epsfig{file=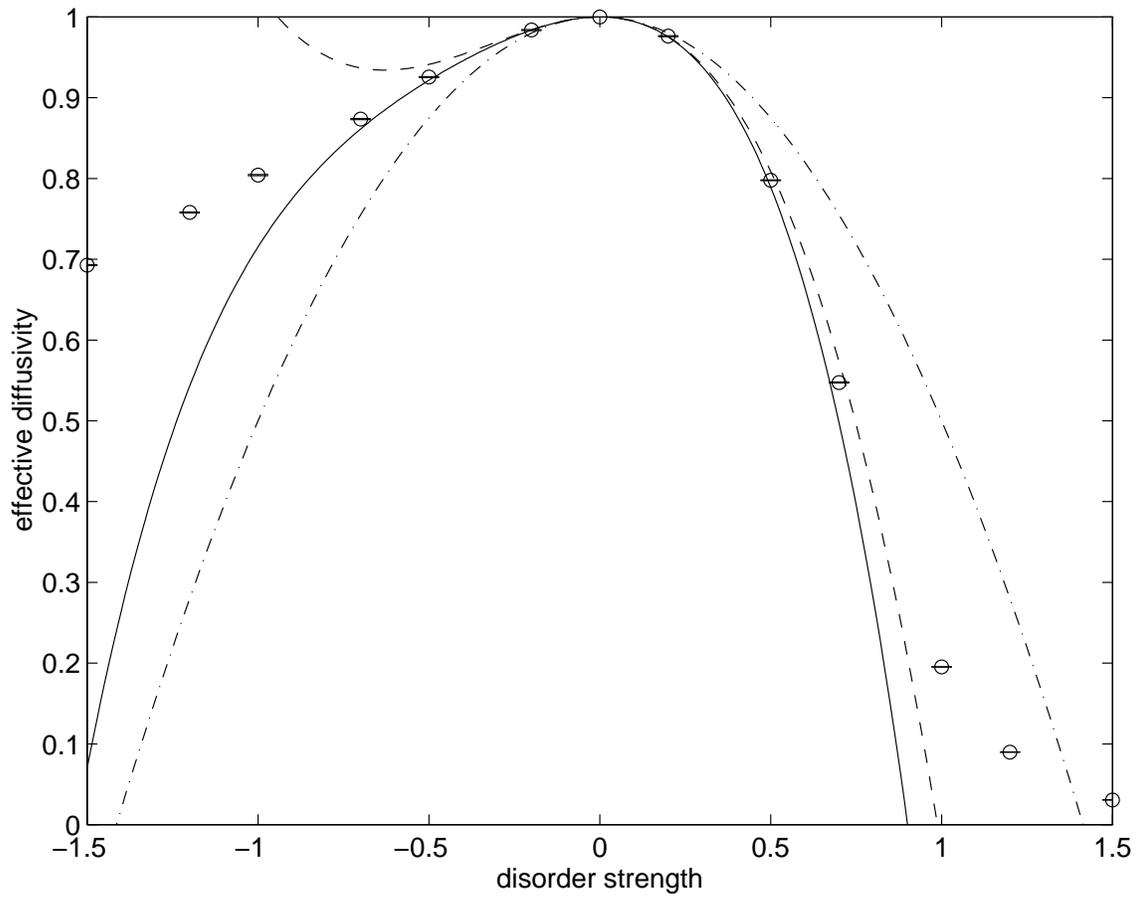,height=120mm}
\end{center}
\caption{\label{f5}$\k_{e}$ versus $\lambda_{0}$ assuming $\k_{0}=1$. The simulation
data are shown (o) to be compared with the prediction of two-loop (dot-dashed),
three-loop (dashed) and four-loop (solid) perturbation theory} 
\end{figure}

This last contribution is the first of the odd power terms in the expansion 
that are responsible for the asymmetry of $\k_{e}$ under change of sign of 
the coupling $\lambda_{0}$~. Its presence is a direct confirmation of the non-Gaussian property 
of the statistics of the velocity field. However, truncating the power series 
at $O(\lambda_{0}^{3})$ causes $\k_{e}$ to increase for larger negative values 
of $\lambda_{0}$, as is shown in figure \ref{f5}. This unphysical feature is obviously 
an artifact of perturbation theory and can be circumvented by including the fouth-order 
term in the perturbation expansion. The diagrams corresponding to this order are 
shown in figure \ref{f4} and give the following contributions:

\begin{eqnarray}
\Sigma^{(4a)}({\bf k})&=&\frac{1}{4}\frac{\lambda_{0}^{4}}{\k_{0}^{3}}\int
\frac{d^{3}{\bf q}}{(2\pi)^{3}}\frac{d^{3}{\bf q'}}{(2\pi)^{3}}\frac{d^{3}{\bf p}}{(2\pi)^{3}}
\frac{d^{3}{\bf p'}}{(2\pi)^{3}}\,\tilde{\Delta}({\bf q})\,\tilde{\Delta}({\bf q'})\,
\tilde{\Delta}({\bf p})\,\tilde{\Delta}({\bf p'})\,{\bf (q\cdot q')}^{2}\,
{\bf (p\cdot p')}^{2}\nonumber \\
& &{\bf k\cdot (p+p')\,(k-p-p')\cdot (q+q')}\nonumber \\
& &\cdot\,\frac{\bf (k-p-p'-q-q')\cdot (p+p')\,(k-q-q')\cdot (q+q')}
{{\bf (k-p-p')}^{2}\,{\bf (k-p-p'-q-q')}^{2}\,{\bf (k-q-q')}^{2}}~~,
\end{eqnarray}
\begin{eqnarray}
\Sigma^{(4b)}({\bf k})&=&\frac{1}{4}\frac{\lambda_{0}^{4}}{\k_{0}^{3}}
\int\frac{d^{3}{\bf q}}{(2\pi)^{3}}\frac{d^{3}{\bf q'}}{(2\pi)^{3}}\frac{d^{3}{\bf p}}{(2\pi)^{3}}
\frac{d^{3}{\bf p'}}{(2\pi)^{3}}\,\tilde{\Delta}({\bf q})\,\tilde{\Delta}({\bf q'})\,
\tilde{\Delta}({\bf p})\,\tilde{\Delta}({\bf p'})\,{\bf (q\cdot q')}^{2}\,
{\bf (p\cdot p')}^{2}\nonumber \\
& &{\bf k\cdot (p+p')\,(k-p-p')\cdot (q+q')}\nonumber \\
& &\cdot\,\frac{\bf (k-p-p'-q-q')\cdot (q+q')\,(k-p-p')\cdot (p+p')}
{{\bf (k-p-p')}^{4}\,{\bf (k-p-p'-q-q')}^{2}}~~,
\end{eqnarray}
\begin{eqnarray}
\Sigma^{(4c)}({\bf k})&=&-\frac{\lambda_{0}^{4}}{\k_{0}^{3}}\int\frac{d^{3}{\bf q}}{(2\pi)^{3}}
\frac{d^{3}{\bf q'}}{(2\pi)^{3}}\frac{d^{3}{\bf p}}{(2\pi)^{3}}
\frac{d^{3}{\bf p'}}{(2\pi)^{3}}\,\tilde{\Delta}({\bf q})\,
\tilde{\Delta}({\bf q'})\,\tilde{\Delta}({\bf p})\,\tilde{\Delta}({\bf p'})\,
{\bf q\cdot q'\,p\cdot p'}\nonumber\\
& &{\bf p'\cdot q\,p\cdot q'\,k\cdot (p+p')\,(k-p-p')\cdot (q-p')}\nonumber\\
& &\cdot\,\frac{\bf (k-p-q)\cdot (q'-q)\,(k-p-q')\cdot (p+q')}
{{\bf (k-p-p')}^{2}\,{\bf (k-p-q)}^{2}\,{\bf (k-p-q')}^{2}}~~,
\end{eqnarray}
\begin{eqnarray}
\Sigma^{(4d)}({\bf k})&=&-\frac{\lambda_{0}^{4}}{\k_{0}^{3}}
\int\frac{d^{3}{\bf q}}{(2\pi)^{3}}\frac{d^{3}{\bf q'}}{(2\pi)^{3}}\frac{d^{3}{\bf p}}{(2\pi)^{3}}
\frac{d^{3}{\bf p'}}{(2\pi)^{3}}\,\tilde{\Delta}({\bf q})\,
\tilde{\Delta}({\bf q'})\,\tilde{\Delta}({\bf p})\,
\tilde{\Delta}({\bf p'})\,{\bf q\cdot q'\,p\cdot p'}\nonumber\\
& &{\bf p'\cdot q\,p\cdot q'\,k\cdot (p+p')\,(k-p-p')\cdot (q-p')}\nonumber\\
& &\cdot\,\frac{\bf (k-p-q)\cdot (q'-p)\,(k-q-q')\cdot (q+q')}{{\bf (k-p-p')}^{2}\,
{\bf (k-p-q)}^{2}\,{\bf (k-q-q')}^{2}}~~,
\end{eqnarray}
\begin{eqnarray}
\Sigma^{(4e)}({\bf k})&=&\frac{\lambda_{0}^{4}}{\k_{0}^{3}}\int\frac{d^{3}{\bf q}}{(2\pi)^{3}}
\frac{d^{3}{\bf q'}}{(2\pi)^{3}}\frac{d^{3}{\bf p}}{(2\pi)^{3}}\frac{d^{3}{\bf p'}}{(2\pi)^{3}}\,
\tilde{\Delta}({\bf q})\,\tilde{\Delta}({\bf q'})\,\tilde{\Delta}({\bf p})\,
\tilde{\Delta}({\bf p'})\,{\bf q\cdot q'\,p\cdot p'}\nonumber\\
& &{\bf p'\cdot q'\,p\cdot q\,k\cdot (p+p')\,(k-p-p')\cdot (q+q')}\nonumber\\
& &\cdot\,\frac{\bf (k-q-q'-p-p')\cdot (q'+p')\,(k-q-p)\cdot (q+p)}
{{\bf (k-p-p')}^{2}\,{\bf (k-q-q'-p-p')}^{2}\,{\bf (k-q-p)}^{2}}~~.
\end{eqnarray}
The above expressions can be simplified using the same sort of manipulations as 
in the three-loop case. The first two are calculated analytically to $O({\bf k}^{2})$ 
with the result
\begin{equation}
\Sigma^{(4,a+b)}({\bf k})=-\frac{1}{8}\frac{\lambda_{0}^{4}}{\k_{0}^{3}}\,{\bf k}^{2}
{\left[\frac{1}{3}\int\frac{d^{3}{\bf q}}{(2\pi)^{3}}\,{\bf q}^{2}\,
\tilde{\Delta}({\bf q})\right]}^{4}=-\frac{1}{8}\frac{\lambda_{0}^{4}}{\k_{0}^{3}}\,{\bf k}^{2}~~.
\end{equation}
The remaining ones lead, again, to a dominant contribution that can be calculated 
analytically plus smaller pieces which are evaluated numerically to yield
\begin{equation}
\Sigma^{(4,c+d+e)}({\bf k})=\frac{\lambda_{0}^{4}}{\k_{0}^{3}}\,{\bf k}^{2}\,(\frac{1}{2}-0.06)~~.
\end{equation}

The outcome for $\k_{e}$ to $O(\lambda_{0}^{4})$ is then
\begin{equation}
\k_{e}=\k_{0}\left\{ 1-\frac{1}{2}\frac{\lambda_{0}^{2}}{\k_{0}^{2}}
-0.530375\frac{\lambda_{0}^{3}}{\k_{0}^{3}}-0.315\frac{\lambda_{0}^{4}}{\k_{0}^{4}}\right\}~~.
\end{equation}

The results for $\k_{e}$ at two, three and four loops are shown plotted in 
figure \ref{f5} for the range $-1.5<\lambda_{0}<1.5$ and $\k_{0}=1$. Clearly, the 
four-loop perturbative calculation is successful in surmounting the 
difficulties of the three-loop case while encoding the deviations from 
Gaussian behaviour. The latter, are translated into a fast decay of $\k_{e}$ 
for positive values of $\lambda_{0}$ whereas it decreases more slowly for the negative values.

\section*{3 Numerical Simulation of Drift and Diffusivity}

To simulate the evolution of the scalar field $P({\bf x},t)$ we integrate 
numerically the stochastic equation for the evolution of a particle with 
path ${\bf x}(t)$ given by equation (\ref{sde}). The resulting probability distribution 
for the particle position ${\bf x}(t)$ is then $P({\bf x},t)$ with the initial 
condition $P({\bf x},0)=\delta({\bf x})$.

The discrete form of equation (\ref{sde}) suitable for numerical integration is
\begin{equation}
{\bf x}_{n+1}-{\bf x}_{n}={\bf u}({\bf x}_{n})\Delta t+(2\k_{0}\Delta t)^{1/2}{\bf \xi}_{n}~~,
\end{equation}
where ${\bf \xi}_{n}$ is a Gaussian random three-vector of zero mean and 
unit variance for each component. This equation models equation (\ref{sde}) correctly 
to $O(\Delta t)$ but the details of a third-order Runge-Kutta scheme correct 
to $O(\Delta t^{3})$ are given in \cite{DrDuH}. We use this third-order scheme 
in our numerical simulation.

The realizations of the random field $\phi({\bf x})$ are constructed in the 
usual way \cite{Kraich,DrDuH,DrH}. We set
\begin{equation}
\phi({\bf x})=\left(\frac{2}{N}\right)^{1/2}\sum_{n=1}^{N} 
\cos({\bf k}_{n}\cdot {\bf x}+{\bf \epsilon}_{n})~~,
\end{equation}
where the vector ${\bf \epsilon}_{n}$ is distributed uniformly over the unit 
sphere and the wavevector ${\bf k}_{n}$ is distributed according to the distribution
\begin{equation}
P({\bf k})=\frac{1}{(2\pi)^{3/2}}\,e^{-{\bf k}^{2}/2}~~.
\end{equation}
For $N$ sufficiently large the central limit theorem guarantees 
that $\phi({\bf x})$ is Gaussian up to $O(1/N)$ corrections.

The effective diffusivity is computed, for a realization of the velocity 
field, from the ensemble of paths by
\begin{eqnarray}
{\langle {\bf x}(t)\cdot {\bf x}(t)\rangle}_{paths}&
=&\lim_{M\rightarrow\infty}\frac{1}{M}\sum_{a=1}^{M} 
{\bf x}^{a}(t)\cdot {\bf x}^{a}(t)\nonumber\\
&=&6\k_{e}t+O(1)\;\;as\;\;t\rightarrow\infty~~.
\end{eqnarray}

To measure the effective drift, $\lambda_{e}$, we add a constant drift 
term to equation (4). In appropriate units, it is given by
\begin{equation}
{\bf u'}=\lambda_{0}{\bf g}~~,
\end{equation}
where ${\bf g}$ is a uniform gradient field. Assuming that the 
latter lies along the $x$-axis, for a realization of the velocity 
field, $\lambda_{e}$ is computed according to
\begin{eqnarray}
{\langle x(t)\rangle}_{paths}&=&\lim_{M\rightarrow\infty}
\frac{1}{M}\sum_{a=1}^{M} x^{a}(t)\nonumber\\
&=&\lambda_{e}gt+O(1)\;\;as\;\;t\rightarrow\infty~~.
\end{eqnarray}

In practise, the number of field realizations and $M$ are finite but 
large enough to give an estimate of $\k_{e}$ and $\lambda_{e}$ with reasonable error. 
In addition the simulation must be carried to values of $t$ large enough to ensure 
that measurements are being performed in the asymptotic regime controlled by the long-range 
or ``renormalized'' parameters. This is tested by ensuring 
that the estimates for $\k_{e}$ and $\lambda_{e}$ are independent of the range of $t$ 
used evaluate them, within statistical errors. In our simulation we tracked the trajectories of 400 
particles in each of 1280 realizations of the velocity field, each of them containing 
128 modes. The simulation was run for a total of 8000 (32000) time steps of 
length 0.025 (0.0125) for the smaller (larger) absolute values of the bare 
coupling $\lambda_{0}$. For larger values of $\lambda_{0}$ and because of 
time limitations the number of paths followed was reduced to a minimum of a hundred.

The results from the perturbation theory calculation of $\k_{e}$ against results 
from the simulation are shown in figure \ref{f5} for the range $-1.5<\lambda_{0}<1.5$ 
and assuming $\k_{0}=1$. The simulation confirms the pronounced asymmetry 
of $\k_{e}$ as a function of $\lambda_{0}$, exhibiting a fast (slow) decay 
for positive (negative) values of $\lambda_{0}$. The four-loop order results 
compare well with the simulation outcome in the range $-0.7<\lambda_{0}<0.7$. 
As expected, the four-loop perturbative calculation encodes the relevant 
qualitative features of the flow and is effective in the region of small 
disorder as opposed to the two-loop and three-loop case. 
\begin{table}
\begin{center}
\begin{tabular}{|c|c|c|c|}
\hline
$\lambda_{0}$&g&$\k_{e}$&$(\lambda_{e}/\lambda_{0})$\\
\hline
\hline
-1.5&0.05&0.69276(60)&0.6946(9)\\
\hline
-1.2&0.05&0.75797(66)&0.7583(12)\\
\hline
-1.0&0.05&0.80412(69)&0.8032(15)\\
\hline
-0.7&0.05&0.87360(75)&0.8753(31)\\
\hline
-0.5&0.05&0.92560(56)&0.9213(32)\\
\hline
-0.2&0.125&0.98386(59)&0.9803(33)\\
\hline
0.2&0.125&0.97632(59)&0.9790(33)\\
\hline
0.5&0.05&0.79788(49)&0.7983(30)\\
\hline
0.7&0.05&0.54744(52)&0.5458(25)\\
\hline
1.0&0.05&0.19537(25)&0.1939(8)\\
\hline
1.2&0.05&0.08979(15)&0.0895(5)\\
\hline
1.5&0.05&0.03059(7)&0.0296(3)\\
\hline
\end{tabular}
\end{center}
\caption{Measurements of $\k_{e}$ and $\lambda_{e}$ for various values of the 
disorder parameter $\lambda_{0}$.}
\end{table}

In table 1 we show the measurements of both $\k_{e}$ and $\lambda_{e}$ 
over a range of values of the disorder parameter (again, we take $\k_{0}=1$). 
The results clearly show that the equality of the two renormalized factors 
is well maintained throughout with only slight discrepancies some cases
due to systematic errors. For the higher values of the disorder parameter 
another possible source of error is that the value of the drift parameter has become 
so large that $O(g^{2})$ effects are influencing the values of the measured 
quantities. Nevertheless, we are confident that the simulation supports the 
conclusion that in gradient flow, irrespective of the precise nature of the 
statistics of the velocity field, the drift and diffusivity parameters are 
renormalized in the same way if we start from a situation where the corresponding 
microscopic quantities are proportional. This result confirms the 
theoretical prediction previously obtained in \cite{DDH5} 
using a theoretical approach developed for the continuum from a method
due to Derrida \cite{Bouch,Derr,BouchLeD}.

\section*{4 Conclusions}

In this paper we have studied the motion of a neutral molecule in a random 
gradient flow. The suggested physical mechanism giving rise to the 
local drift involves the interaction of the local electric field
(presumed to have Gaussian statistics) with the field-induced dipole
moment of the molecule. The resulting drift field therefore does {\it not} have
Gaussian statistics. That such a physically plausible model can give rise, 
in a natural way, to a drift field with
non-Gaussian statistics, is itself interesting.

The increased complexity of the model is no barrier to the formulation of
a perturbative calculation of its Green's functions and effective parameters.
The non-Gaussian character of the drift field statistics shows up first
at three loop order. It results in a contribution to the effective diffusivity
that is not symmetric under a change of sign of the coupling. This asymmetry
is very clear in both the perturbation theory and numerical results exhibited
in figure \ref{f5}~. The same figure also exhibits the results of the numerical simulation.
It shows that the inclusion of the four loop
terms allows the perturbation series to give a reasonably good account of the effective
diffusivity for the coupling parameter in the range $-0.7<\lambda_{0}<0.7$~.

The complexity of the non-Gaussian model, however, has so far prevented, a satisfactory
formulation of the kinds of self-consistent perturbative calculation or 
renormalization group calculation that were rather successful for the
Gaussian model \cite{Krav1,Krav2,Deem,DeanTh,DDH1}. In both cases the problem centres round
the new types of vertices, not present in the original perturbation
theory scheme, that are {\it induced} by the loop contributions of perturbation theory. 
The formulation of effective calculational schemes of this kind remains 
a goal of great interest since, as is clear from figure \ref{f5}, 
low order perturbation theory is inadequate
for situations of large disorder. Further investigations are in progress.

In addition to the effective diffusivity the effective drift
parameter, which controls the response of the molecules to an 
externally applied constant gradient field, is also a significant
physical quantity. The success of the renormalization group calculation in the Gaussian model
was in part due to the fact that it respected the Einstein relation, namely
that the effective diffusivity and drift parameter are renormalized by the same
factor from their molecular values. The Einstein relation was also demonstrated
in low order perturbation theory. In fact the Einstein relation
was later shown to hold quite generally, independently of the statistical
properties of the random medium, provided the molecular drift and diffusivity
tensors were proportional to one another \cite{DDH5}:
the Gaussian character of the  model is not relelvant.
It is therefore important that
we have been able to give numerical confirmation (see table 1) 
of the validity of the Einstein relation 
between effective drift and diffusivity in the non-Gaussian model 
investigated in this paper. 

The introduction of directional effects into the drift and diffusivity as well
as the statistics of the drift field, are further problems of great
interest but considerably increased complexity.
\section*{Acknowledgments}

The calculations were performed on the Hitachi SR2100 located at the University of 
Cambridge High Performance Computing Facility. C A da Silva Santos wishes to 
thank JNICT-Progama PRAXIS XXI for financial support under grant BD/3126/94.

\newpage
\end{document}